\documentclass{nsf_proposal}

\usepackage{times}
\usepackage{amsmath, amsthm, amssymb}
\usepackage{natbib}
\setcitestyle{numbers}
\newtheoremstyle{taskstyle}
{.05in}
{0in}
{\itshape}
{}
{\bfseries}
{.}
{.1in}
{}
\theoremstyle{taskstyle}
\newtheorem{tstyle}{Challenge}
\usepackage{titlesec}
\titlespacing\section{0pt}{2pt plus 0pt minus 2pt}{0pt plus 0pt minus 2pt}
\begin{document}
	
		\thispagestyle{empty}

\newcommand{\hs}{0.2in}
\noindent

\clearpage
\pagenumbering{arabic} 
\date{}

\title{\vspace{-.8in}
	Challenges Towards Deploying Data Intensive Scientific Applications on Extreme Heterogeneity Supercomputers\vspace{-.2in}}
\author{
Hang Liu\textsuperscript{$\ast$}\hspace{\hs} 
Yufei Ding\textsuperscript{$\ddagger$} \hspace{\hs} 
Da Zheng\textsuperscript{$\dagger$} \hspace{\hs}
Seung Woo Son\textsuperscript{$\ast$}  \hspace{\hs} 
Da Yan\textsuperscript{$\diamond$}
\\
$\ast$: UMass Lowell\hspace{\hs} 
$\ddagger$: UCSB\hspace{\hs}
$\dagger$: Unaffiliated\hspace{\hs} 
$\diamond$: UAB
\\
{\tt Hang\_Liu@uml.edu}
\vspace{-.25in}
}
\maketitle




\begin{abstract}

Shrinking transistors, which powered the advancement of computing in the past half century, has stalled due to power wall; now extreme heterogeneity is promised to be the next driving force to feed the needs of ever-increasingly diverse scientific domains. \textit{To unlock the potentials of such supercomputers, we identify eight potential challenges in three categories}: First, one needs fast data movement since extreme heterogeneity will inevitably complicate the communication circuits -- thus hampering the data movement. Second, we need to intelligently schedule suitable hardware for corresponding applications/stages. Third, we have to lower the programming complexity in order to encourage the adoption of heterogeneous computing.

\end{abstract}

\section{Introduction}

Extreme heterogeneity is the result of using multiple types of processors, accelerators and memory/storage in a single computing platform, which is different from traditional Top 500 supercomputers~\cite{top2017www} that install at most two types of processors, one types of accelerator and memory/storage according to their specifications (e.g., Titan, Sequoia, Trinity and Cori). Clearly, higher (or extreme) heterogeneity supercomputing will enable the support of a variety of application workflows and meet the needs of increasingly diverse scientific domains. 


This is especially true for data intensive scientific applications, such as, High-Performance Conjugate Gradient (HPCG)~\cite{hpcg2017bench,mittal2016computational} and High-Performance Data Analytics (HPDA), each of which comprise of a diverse collection of operations. In particular, HPCG involves a sequence of repeated Basic Linear Algebra Subprograms (BLAS) operations, e.g., \textsf{SpMV, xaxpy, xdot, xcopy and xxdot} and dissimilar BLAS operations that prefer different processors and accelerators. For instance, \textsf{SpMV} prefers GPUs while \textsf{xaxpy and xxdot} may benefit more on many-core CPUs and \textsf{xdot and xcopy} should favor multi-core CPUs. HPDA stands for a collection of data mining, graph computing and machine learning algorithms~\cite{liu2015enterprise,da2015flashgraph,ding2015yinyang}. These applications typically process a daunting volume of data and generate nontrivial amount of algorithmic metadata with random access patterns on both data structures. As such, a richer hierarchy of memory/storage from extreme heterogeneity will provide users more options to cache the data with respect to its size and reuse frequency.



Putting hardware together is important, a system that can sufficiently and intelligently extract the potentials of corresponding hardware is even more crucial. Briefly, we envision there is a need to address challenges along the following three directions to unleash the capabilities of extreme heterogeneity computing. 
First, one needs to provide fast data movement from storage media to processors since installing more devices together will inevitably complicate the communication circuit, and thus hamper the data movement. Second, intelligently suggesting the most suitable hardware for corresponding application (stages) with dynamic adjustment is essential. Third, we have to lower the programming entry bar for both expert and non-expert users in order to encourage the adoption of heterogeneous computing.


\section{Data Movement}
Despite deeper memory/storage hierarchy provides users the opportunity of caching data at various hierarchies according to their reuse frequencies, it may also elongate the distance between data and processing units. We admit data movement is yet a significant bottleneck for contemporary computing nodes because they are in the form of \textit{modest heterogeneity}. However, such adversarial effects will deteriorate rapidly as the system heterogeneity climbs towards extremum. For instance, recent study~\cite{zhang2015numa} demonstrates that the cross non-uniform memory access (NUMA) node memory access throughput drops from 20\% on a two-socket (i.e., Intel Xeon E5-2683 v3) processor to 6.85$\times$ of an eight-socket (i.e., Intel Xeon E7-8850 v3) processor.



Unfortunately, contemporary Operating Systems (OSes)~\cite{bovet2005understanding,liu2017graphene} pay seldom attention on the potential drawbacks of heterogeneities. Particularly, they disregard the differences of circuit distance between processor, accelerator and storage media caused by heterogeneity, reflected as they follow round robin fashion to assign processors and accelerators to applications, as well as storage media. In addition, even computation only happens on accelerators, they still require the application to copy the data from storage media to CPU memory before further sending the data to accelerators. Beyond that, existing pagecache relies on two-layer FIFO queue policy to cache all data that is fetched from disk to CPU memory. Assuming the should be cached data fails to satisfy two-layer FIFO queue policy, which is most likely the case in HPDA, current pagecache may waste both memory space and time to cache and query the data~\cite{van2015di}. As such, we advocate to address the following challenges in this field.
\begin{tstyle}
Need to configure affinity across storage, memory, processor and accelerator.
\end{tstyle}

\begin{tstyle}
Need to introduce direct data access from accelerator to storage.
\end{tstyle}

\begin{tstyle}
Need user controllable page cache with a variety of caching policies.
\end{tstyle}

\vspace{.2in}

\section{Resource Management}


Albeit with certain drawbacks, we envision the future extreme heterogeneity supercomputer can be a strong suit for today's increasingly irregular scientific applications 
if with intelligent resource management. We will illustrate the computation irregularity of scientific applications with cardiac simulation~\cite{mittal2016computational} which simulates the blood flow patterns in left ventricle. Particularly, this process needs to, in an interleaved manner, simulate muscle systolic and diastolic, as well as blood flow entering and exiting the heart as time goes by. In this application, muscle simulation involves tremendous data access thus prefers many-core CPUs, blood simulation expects more benefits in accelerators while the initializations of both stages perform better at multi-core CPUs.



Similarly, various hierarchies of memory/storage, which have different capacities, throughputs and latencies, if used effectively, can help enhance performance, reduce cost and save energy. For instance, Neurodata storage system \cite{neurodata2013open} lives in the cloud. It uses S3 for the backend storage and uses memory and SSDs for caching. To accelerate different types of queries, such as data ingestion and data analytics, this system maintains separate caching layers, which further increase the effectiveness of resource utilization and performance in the system.
However, neither of the above benefits will be possible for scientific computing if we do not have a toolkit that can intelligently select correct hardware for distinct datasets and application. Consequently, we envision the following resource management challenges in this category.

\begin{tstyle}
Need a judicious processor selection based on the execution features of various applications (stages).
\end{tstyle}

\begin{tstyle}
Need an intelligent data distribution across HDD, SSD, NVMe, fast and slow memory components.
\end{tstyle}

\section{Compiler Framework}
Demand for expertise in various programming models and a spectrum of hardware architectures should be relieved from application-domain expects, otherwise, it may discourage them from deploying their applications in a heterogeneous supercomputing system as exhaustively developing source codes for all types of hardware is both time consuming and labor intensive. To utilize a contemporary modest heterogeneity supercomputer with CPUs, GPUs, and FPGAs, for instance, users need to learn MPI/OpenMP, CUDA/OpenCL and Verilog/VHDL.

{

One common remedy is to develop functional libraries for each of these processors. Upon these libraries, a set of unified APIs can then be designed to alleviate the process-specific programming efforts. This turns out to be a reasonable attempt, and it has been used extensively in the deep learning domain. 
For example, Tensorflow provides APIs in both C++ and python, with which users can deploy their program in CPUs or GPUs without learning to program in MPI/OpenMP or CUDA/OpenCL. 


However, such library-API methodology suffers from two major drawbacks. First, tuning a special-purpose library itself could take over years 
from even domain experts. For instance, cuDNN for GPU, as a specialized version of the dense linear algebra libraries (BLAS) for CPU, was tuned by a team of expert programmers in NVIDIA and was not released until two years of extensive tunning. 
Second, high-level APIs often implicitly sacrifice performance for abstractions. 
It is known that cross-layer and other large-scope optimizations are missing in Tensorflow and other library-API based frameworks, as those supplied APIs are at the layer level (e.g., convolution layer, pooling layer, fully-connected layer).
Consequently, a program written with these APIs will, by default, suffer from the loss of  of those optimizations.

To resolve these problems, we suggest building a more powerful compiler framework which will tackle the following challenges.

 
\begin{tstyle}
Need a compiler to translate source code from some high-level programming languages, e.g., python to a process-specific language, e.g., CUDA.
\end{tstyle}

\begin{tstyle}
The compiler can apply a good set of optimizations to the translated programs, including traditional optimizations like loop titling and fusion, as well as some predefined/user-specified domain-specific optimizations.
\end{tstyle}

\begin{tstyle}
Need a runtime that examines the data flow of the program and other dynamic information for enabling a larger set of optimizations dynamically, including those cross-layer and other large-scope optimizations.
\end{tstyle}
}

\setlength{\bibsep}{0.0pt}

\bibliographystyle{plain}
{
\scriptsize
\bibliography{main}

\begin{thebibliography}{10}

\bibitem{bovet2005understanding}
Daniel~P Bovet and Marco Cesati.
\newblock {\em Understanding the Linux Kernel: from I/O ports to process
  management}.
\newblock O'Reilly Media, Inc., 2005.

\bibitem{neurodata2013open}
Randal Burns et~al.
\newblock The open connectome project data cluster: Scalable analysis and
  vision for high-throughput neuroscience.
\newblock In {\em SSDBM}, 2013.

\bibitem{ding2015yinyang}
Yufei Ding et~al.
\newblock Yinyang k-means: A drop-in replacement of the classic k-means with
  consistent speedup.
\newblock In {\em ICML}, 2015.

\bibitem{hpcg2017bench}
{High-Performance Conjugate Gradient (HPCG) Benchmark}.
\newblock {http://www.hpcg-benchmark.org/}, 2017.

\bibitem{liu2015enterprise}
Hang Liu et~al.
\newblock Enterprise: Breadth-first graph traversal on gpus.
\newblock In {\em SC}, 2015.

\bibitem{liu2017graphene}
Hang Liu and H~Howie Huang.
\newblock Graphene: Fine-grained io management for graph computing.
\newblock In {\em FAST}, 2017.

\bibitem{mittal2016computational}
Rajat Mittal, Jung~Hee Seo, Hang Liu, et~al.
\newblock Computational modeling of cardiac hemodynamics: current status and
  future outlook.
\newblock {\em JCP}, 2016.

\bibitem{top2017www}
{Top 500 November Ranking}.
\newblock {https://www.top500.org/lists/2017/11/}, 2017.

\bibitem{van2015di}
Brian Van~Essen, Henry Hsieh, Sasha Ames, Roger Pearce, and Maya Gokhale.
\newblock Di-mmap -- a scalable memory-map runtime for out-of-core
  data-intensive applications.
\newblock {\em Cluster Computing}, 2015.

\bibitem{zhang2015numa}
Kaiyuan Zhang, Rong Chen, and Haibo Chen.
\newblock Numa-aware graph-structured analytics.
\newblock In {\em PPoPP}, 2015.

\bibitem{da2015flashgraph}
Da~Zheng et~al.
\newblock Flashgraph: Processing billion-node graphs on an array of commodity
  ssds.
\newblock In {\em FAST}, 2015.

\end{thebibliography}
}

\end{document}